\documentclass[footinbib,amsmath,amssymb,superscriptaddress,prl,reprint]{revtex4-1}
\usepackage[utf8]{inputenc}
\usepackage{graphicx}
\usepackage{xcolor}
\usepackage{hyperref}
\usepackage{comment}

\usepackage{notes2bib}

\usepackage[version=3]{mhchem}
\usepackage{siunitx}

\newcommand{\rs}[1]{_{\rm #1}} 

\usepackage{printlen}
\uselengthunit{in}

\newcommand{\FigCrys}{Supplementary Figure 1}

\newcommand{\FigAging}{Supplementary Figure 4}
\newcommand{\FigBrokenBonds}{Supplementary Figure 5}

\begin{document}
\title{Active glass: ergodicity breaking dramatically affects response to self-propulsion}
\author{Natsuda Klongvessa}
\author{F\'elix Ginot}
\altaffiliation[Now in: ]{Fachbereich Physik, Universit\"at Konstanz, Universit\"atsstrasse 10, 78464 Konstanz, Germany}
\author{Christophe Ybert}
\author{C\'ecile Cottin-Bizonne}
\author{Mathieu Leocmach}
\email[\href{https://twitter.com/LamSonLeoc}{@LamSonLeoc} or ]{mathieu.leocmach@univ-lyon1.fr}
\affiliation{Universit\'e de Lyon, Universit\'e Claude Bernard Lyon 1, CNRS, Institut Lumi\`ere Mati\`ere, F-69622, VILLEURBANNE, France}

\begin{abstract}

We study experimentally the response of a dense sediment of Brownian particles to self-propulsion. We observe that the ergodic supercooled liquid relaxation is monotonically enhanced by activity. By contrast the nonergodic glass shows an order of magnitude slowdown at low activities with respect to passive case, followed by fluidization at higher activities.
Our results contrast with theoretical predictions of the ergodic approach to glass transition summing up to a shift of the glass line.
We propose that nonmonotonicity is due to competing effects of activity: (i) extra energy that helps breaking cages (ii) directionality that hinders cage exploration. 
We call it ``Deadlock from the Emergence of Active Directionality'' (DEAD).
It suggests further theoretical works should include thermal motion.


\end{abstract}

\maketitle

A supercooled liquid is obtained when a system is cooled down, or compressed, beyond its freezing temperature while avoiding crystallization. This metastable state displays slow dynamics but remains ergodic. 
As the system is further cooled down or compressed, its dynamics slows  down by orders of magnitude until the system becomes nonergodic, which means that it can explore only a small part of its potential energy landscape. 
It is an amorphous solid called a glass. Our understanding of this fundamental state of matter has tremendously progressed in the last decades~\cite{Cavagna2009,CharbonneauReview2017}. 
Studying the glass transition under nonequilibrium conditions helps us define what are general properties of glassy systems and their emergent behaviors when they are driven out-of-equilibrium. This is where the field of active matter, which emerged as a new frontier of science, meets glassy physics. In the past years, the behavior of assemblies of self-propelled objects stepped up from a mere zoological curiosity to a flourishing field of nonequilibrium physics. 
Rather dilute assemblies of active particles have been studied extensively by experiments and numerical simulations~\cite{Ballerini2008, Deseigne2010,Theur2012,Bricard2013, Nishi2015, Bechinger_rmp-2016, Marchetti2013}.  
Exploring the full range of densities including ordered phases has been done in some model systems \cite{Digregorio2018, Briand2018, Fily2012, Wysocki2014} but dense amorphous systems remain largely unexplored experimentally.
%
Dense assemblies of self-propelled particles sit at the convergence of active matter and glassy physics, and should constitute a test bed for other such systems as for example biological tissues \cite{bi2016MotilityDrivenGlassJamming,Fodor2018}. 

However, it is still unclear how self-propulsion would influence the glass transition. 
%
%
Numerical studies have found either activity-induced fluidization~\cite{Ni2013b,Berthier2014} or arrest~\cite{Szamel2015,Flenner2016}. 
It was found that the influence of activity could not be captured by a single parameter such as  effective temperature,  but that the persistence time of the propulsion direction played a major role and shifts the position of the glass transition line in nontrivial ways. For example in Ref.~\cite{Berthier2017} glass transition shifts to higher densities with increasing persistence time at low effective temperature, whereas the opposite effect is observed at higher effective temperatures. Besides, Ref.~\cite{Nandi2018} demonstrates that the monotonicity of the glass transition shift depends on the microscopic details of the activity.

Most of the previous numerical studies approached the glass transition from the ergodic supercooled state.
They found that despite a quantitative shift of the glass transition line, the qualitative phenomenology of glassiness remained unchanged~\cite{Berthier2017}.
However, in the present letter, we show experimentally that a 
different, nontrivial phenomenology emerges beyond the glass transition line in the nonergodic glass state.
We study the influence of self-propulsion on a sediment of Brownian particles, in order to access states on both sides of the nonergodic glass transition.
Previous experiments have shown that, in the dilute regime, such active colloids behave like passive colloids with a higher effective temperature~\cite{Ginot2015}.
Indeed from the ergodic side, we observe a monotonic shift of the glass transition line with effective temperature at fixed persistent time.
However in the nonergodic side, we find that low activity levels slow down relaxation of the glass state, followed by a fluidization at higher activity levels, an observation that cannot be rationalized from the concept of effective temperature.
We explain our results by considering how self-propulsion modifies the cage exploration process.
We then discuss how this well-characterized experimental observation fits into the state of our theoretical understanding of active glassy systems.

\begin{figure}
    \centering
    \includegraphics{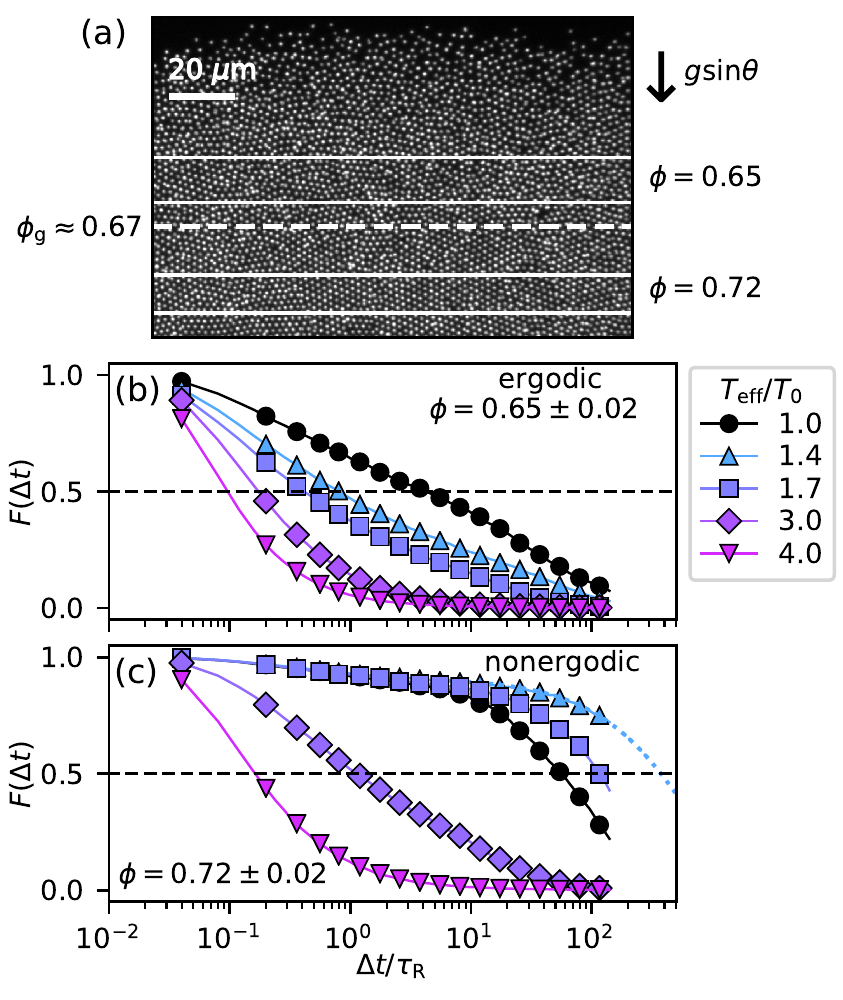}
    \caption{(a) Experimental image of the sediment showing the slicing to get access to different densities. 
    The glass transition density of the passive case is $\phi\rs{g} \approx 0.67$. 
    (b, c) Relaxation function $F(\Delta t)$ for various activity levels at fixed densities $\phi = 0.65 \pm 0.02$ and $0.72 \pm 0.02$, respectively. 
    Horizontal line at 0.5 shows the definition of the relaxation time $\tau$. The dotted curve in (c) is a stretched exponential fit.}
    \label{fig:Ft}
\end{figure}

We study a two-dimensional assembly of gold particles half-coated with platinum~\cite{Howse2007} that behave as soft particles with effective diameter $\sigma_0=\SI{2.2}{\micro\metre}$. Accordingly the hydrodynamic radius we measure is $R_\mathrm{H} \approx \SI{0.94}{\micro\metre}<\sigma_0/2$. We track particles using \textit{trackpy} package~\cite{trackpy2016} and analyze the bond network using \textit{NetworkX} package~\cite{networkX}. We define the area fraction as $\phi=4\varrho/(\pi \sigma_0^2)$, where $\varrho$ is the number density. In the following we normalize distances by $\sigma_0$, and times by rotational Brownian time $\tau_\mathrm{R} = (8\pi\eta R_\mathrm{H}^3)/(k_\mathrm{B}T_0) \approx \SI{5}{\second}$, where $T_0$ is the bath temperature.
%
Upon addition of hydrogen peroxide (\ce{H2O2}), the particles become active and self-propelled~\cite{Paxton2004,Brown2014}. 
In order to access a high density regime, we make in-plane sedimentation which is obtained by tilting the whole set-up with a small angle $\theta \approx \ang{0.1}$ \cite{Ginot2015}. An experimental image is shown in Fig.~\ref{fig:Ft}a. Since the density profile depends on the activity~\cite{Klongvessa2019b}, we parameterize our results by $\phi$. We slice the density profile perpendicularly to gravity so that each slice contains approximately $1000 \pm 100$ particles and has a constant $\phi$ within $0.02$. We then carry analysis on each slice and show the results function of $\phi$ and the activity. 
Note that the polydispersity (10\%) is not enough to prevent local crystallization at high densities (see \FigCrys{} and Ref.~\cite{Nelson1979}). The results presented here exclude crystalline particle and we consider only slices that contains less than 50\% of crystalline particles ($\phi < 0.75 $).


From the sedimentation experiment on passive colloids~\cite{Perrin1909}, the competition between diffusive motion and gravity $g$ results in a density profile that has the Boltzmann form at low enough densities: $\phi(x) \sim \exp[mg x/D_0\mu]$, where $m$ is the buoyancy mass, $x$ is the coordinate in the direction of gravity, $D_0=k\rs{B}T_0/\mu$ is the diffusion coefficient and $\mu= 6\pi\eta R_\mathrm{H}$ is the mobility. Following Refs~\cite{Tailleur2009,Palacci2010}, in the case of self-propelled particles $D_0$ can be replaced by the long time effective diffusion coefficient $D_\mathrm{eff}(\phi \rightarrow 0)$.
For spherical particles undergoing both Brownian and self-propelled motions in 2D but with two degrees of rotational freedom~\cite{Palacci2010,hagen2011BrownianMotionSelfpropelled}, we have $D_\mathrm{eff}(\phi\rightarrow 0) = D_0 + (F_\mathrm{P}/\mu)^2\tau_\mathrm{R}/6$, where $F_\mathrm{P}$ is the magnitude of the propulsion force.

Equivalently $T_0$ can be replaced by an effective temperature such that $k\rs{B}T\rs{eff}\equiv \mu D_\mathrm{eff}(\phi \rightarrow 0)$. This amounts to viewing a dilute active system as ``hot colloids'' with an effective temperature~\cite{Palacci2010}:
\begin{equation}
    \frac{T_\mathrm{eff}}{T_0} = \frac{D_\mathrm{eff}}{D_0} = 1 + \frac{2}{9}\left(\frac{F_\mathrm{P}R_\mathrm{H}}{k_\mathrm{B}T_0}\right)^2.
    \label{eq:TeffPeH}
\end{equation}

In our dense experimental system, we assume that the persistence time is fixed by Brownian rotational diffusion and is thus constant with activity, as observed in dilute conditions~\cite{Klongvessa2019b}. Some of us have shown that this hypothesis is sufficient to explain quantitatively the dynamics of locally closed packed clusters of the same particles~\cite{Ginot2018}. Therefore in the following we characterize activity in every density regimes by $T\rs{eff}/T_0$ measured from the sedimentation profile in the dilute regime.


To characterize the relaxation within a slice, we compute the overlap function~\cite{Flenner2011}, $F(\Delta t)$, which tells us the ratio of particles that have not moved further than $0.3\sigma_0$ during the lag time $\Delta t$.
For instance, in Fig.~\ref{fig:Ft}b and c we show $F(\Delta t)$ at various activities but at two fixed densities $\phi=0.65 \pm 0.02$ and $\phi=0.72 \pm 0.02$, respectively. 
At both densities, the passive case (the black curve) shows two-step relaxation, with almost complete decay of $F(\Delta t)$ within the experimental time. The plateau at the intermediate $\Delta t$ indicates that each particle is trapped by its neighbors. At long times, the system exits the plateau hinting that the particles manage to diffuse away from their original positions. This is a typical glassy behavior.
At high levels of activity ($T\rs{eff}/T_0 = 3.0$ and $4.0$), the plateau disappears and the system completely relaxes. 
At $\phi=0.65 \pm 0.02$, the second relaxation step of $F(\Delta t)$ decreases as $T\rs{eff}$ increases, showing a monotonic response to activity.
By contrast, at $\phi=0.72 \pm 0.02$ the response is nonmonotonic.
%
As we introduce a small amount of activity, the plateau gets longer than the passive case. This surprisingly indicates that the system is less mobile when each particle is weakly self-propelled.
However, when we increase further the activity, the plateau shortens again ($T\rs{eff}/T_0 = 1.7$) and finally disappears at high activity levels ($T\rs{eff}/T_0 = 3.0$ and $4.0$), resulting in decays faster than the passive case.

We call this nonmonotonic behavior of the decay of $F(\Delta t)$ with $T\rs{eff}$ a ``back and forth" behavior. The ``back'' behavior is when the system relaxes slower than the passive case, whereas in the ``forth'' regime the relaxation is enhanced by activity.
We also observe a similar nonmonotonic behavior in the percentage of broken bonds (see \FigBrokenBonds), showing that not only absolute positions but also structure rearrangements are responding in a nonmonotonic way.
The ``forth'' behavior seems rather straightforward: it happens when a particle has enough propulsion force to push its neighbors and move inside the dense phase. 
However the ``back'' behavior is less intuitive and more intriguing. In the following, we will try to understand in which conditions the mobility of the system does depend nonmonotonically on the activity level.

\begin{figure}
    \centering
    \includegraphics{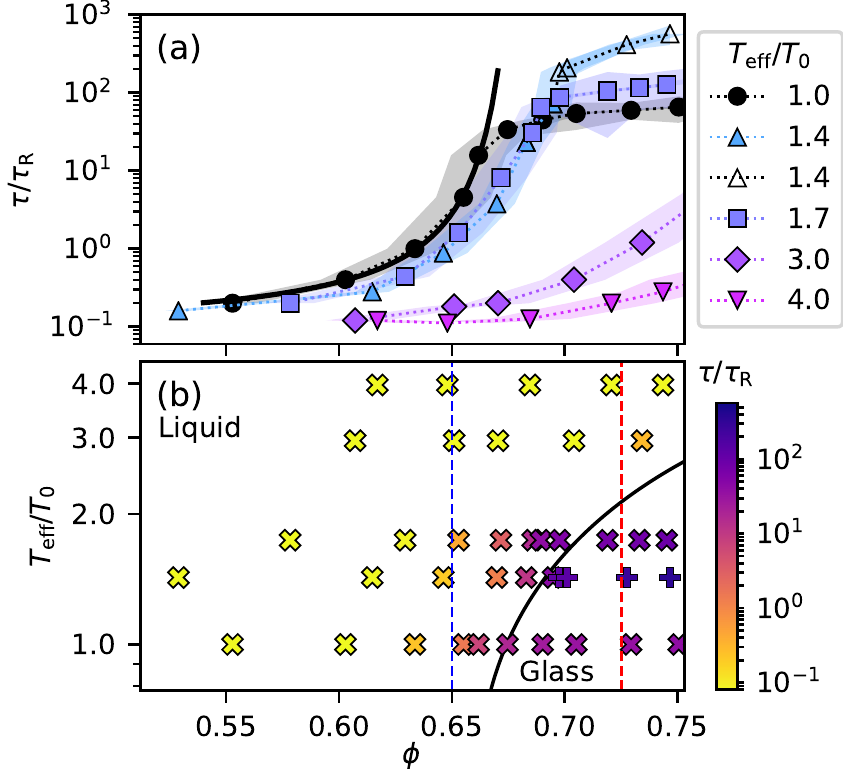}
    \caption{(a) Density dependence of relaxation time $\tau$ at various activities. 
    For $T\rs{eff}/T_0=1.4$, $\tau$ is longer than the maximum lag time at densities higher than 0.70. 
    Open triangles are obtained by extrapolation of $F(\Delta t)$. 
    Transparent areas around curves show uncertainties that come mostly from the uncertainty of area density ($\pm 0.02$) below $\phi_\mathrm{g}$ or the standard deviation of $\tau$ from different sampling above $\phi_\mathrm{g}$. 
    The solid line is the fit $\tau \propto
    \exp{(\frac{A}{(\phi^*/\phi)-1})}$ for $T\rs{eff}/T_0 = 1.0$, where $A \approx 0.19$, and $\phi^* (T_\mathrm{eff}) = 0.69$.
    (b) Dependence on both density and activity of $\tau$, obtained directly from $F(\tau) = 0.5$ (cross symbols), by extrapolation of $F(\Delta t)$ (plus symbols).
    The solid curve is a guide for the eye materializing the glass transition line. 
    Two vertical dashed lines at $\phi = 0.65$ (blue) and $\phi = 0.72$ (red) correspond to the densities in Fig.~\ref{fig:Ft} b and c. }
    \label{fig:diagram} 
\end{figure}


We define the relaxation time $\tau$ when half of the particles have already moved, i.e., $F(\tau) = 0.5$. When $F$ does not reach 0.5 but significantly decays from the plateau, the relaxation time can be estimated by extrapolation (see $T_\mathrm{eff}/T_0=1.4$ in Fig.~\ref{fig:Ft}c).

Fig.~\ref{fig:diagram}a shows how $\tau$ depends on density for various activities. In the passive case (black circles) $\tau$ rises steeply with $\phi$, following a Vogel–Fulcher-like dependence (solid line) until $\phi \approx 0.67$.
Beyond, we observe a saturation of $\tau$, typical of nonergodic glass made of soft particles~\cite{philippeGlassTransitionSoft2018}. In \FigAging{}, we confirm that the relaxation in this regime is waiting-time dependent, symptomatic of the aging of a nonergodic system.
For nonzero activities, the rise of $\tau$ follows the same dependence in $\phi$, shifted towards higher and higher densities (see companion article~\cite{Klongvessa2019b}). We take the first point that deviates from the fit as the operational glass transition packing fraction $\phi\rs{g} (T_\mathrm{eff})$.
We observe that $\phi\rs{g}$ increases monotonically with activity, which is consistent with theoretical expectations for glassy systems with an additional active force at constant persistence time and increasing effective temperature~\cite{Flenner2016,Berthier2017,Nandi2018}.

In general for passive soft particles, the saturation value for $\tau$ only depends on the relaxation time in the dilute limit~\cite{philippeGlassTransitionSoft2018}. By contrast, here we observe a nonmonotonic dependence of the saturation value on $T\rs{eff}/T_0$.
In our lowest nonzero activity ($T\rs{eff}/T_0 = 1.4$, light blue triangles), $F(\Delta t)$ does not decay within our experimental time for all $\phi>0.70$. It implies relaxation times at least an order of magnitude above the saturated $\tau$ in the passive case. Consistently, the last four values of $\tau$ (open triangles) are obtained by the extrapolations of $F$. 
At our second activity ($T\rs{eff}/T_0 = 1.7$, violet squares), we are able to measure a saturated relaxation time about twice longer than in the passive case, that is a decrease with respect to $T\rs{eff}/T_0 = 1.4$. Finally, for higher activities ($T\rs{eff}/T_0 = 3.0$, purple diamonds and $4.0$, pink down triangles), the relaxation time never reaches values beyond the passive case and no saturation is observed within accessible densities.

In Fig.~\ref{fig:diagram}b we map the value of relaxation times on the $(\phi, T\rs{eff}/T_0)$ phase diagram.  
This representation confirms that the glass transition shifts monotonically toward higher densities with increasing activity. The ``forth'' behavior comes from the crossing of glass transition line to the ergodic phase. The ``back'' behavior is observed only when going from zero to nonzero effective temperature in an already nonergodic state. We stress that this nonmonotonic behavior could not be traced by a simple path in the phase diagram.

\begin{figure}
    \centering
    \includegraphics{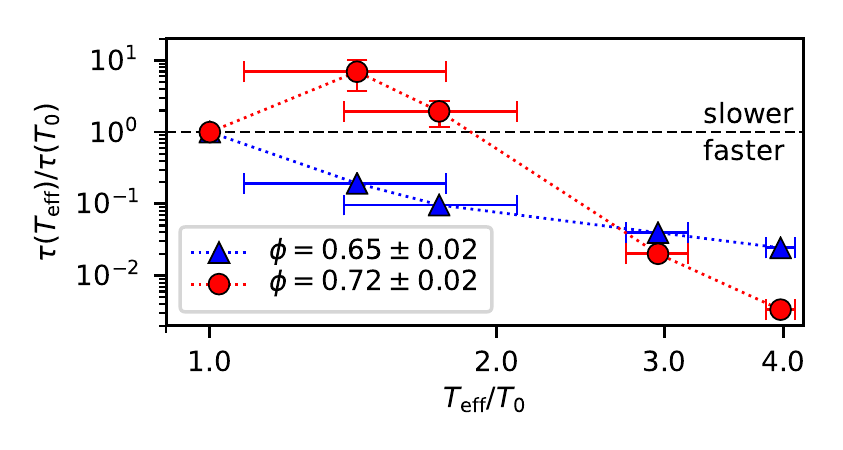}
    \caption{Contrast of activity dependence of $\tau$ between both sides of glass transition. For $\phi=0.72$ (red circles), the first three points are glassy, nonergodic and we observe a nonmonotonic dependence on activity, but not at $\phi=0.65$ (blue triangles) were all points are ergodic. The horizontal dashed line shows $\tau$ in the passive case.} 
    \label{fig:tau_Teff}
\end{figure}

In Fig.~\ref{fig:diagram}b, we draw two vertical lines corresponding to the two densities in Fig.~\ref{fig:Ft}b and c. We then follow both lines starting from $T\rs{eff}/T_0 = 1$ and illustrate the resulting $\tau$ in Fig.~\ref{fig:tau_Teff}.
At $\phi=0.65$ (blue line, triangles), the original passive system is an ergodic supercooled liquid. We observe a monotonic decrease of $\tau$ with increasing $T\rs{eff}$. By contrast, when starting from a passive state that is nonergodic at $\phi = 0.72$ (red line, circles), we observe the nonmonotonic behavior that translates the rise and fall of the saturation level of the relaxation time. $\tau$ increases at low levels of activity and then decreases as the activity increases further. This exemplifies the difference between the respective responses of originally ergodic and nonergodic systems.

We have thus confirmed that the addition of self-propulsion onto a nonergodic glass actually hinders its relaxation. In the following we explain by a scaling argument how a glass of weakly active Brownian particles can relax slower than a glass of passive Brownian particles, and why the transition between the two behaviors is so sudden.

In general, there are two relaxation mechanisms in any dense systems: 
(i) isotropic cooperative motion that involves diffuse broken bonds and 
(ii) collective directed motion that involves no broken bonds inside the correlated region, but at domain boundaries. 
We know that a passive glass relaxes only by the first mechanism~\cite{Cavagna2009}. 
When self-propulsion is introduced, particle motion acquires persistence and the second mechanism is made possible by the particle directed motion. 
At high enough activities, collective motion is dominating: relative positions relax only at very long times (see \FigBrokenBonds) but absolute positions relax faster than in the passive case (Fig.~\ref{fig:Ft}c). This effect is quantified in the companion article~\cite{Klongvessa2019b} within a polycrystal which display the same phenomenology.
What is not obvious is the drop of effectiveness of cooperative movement at the very first nonzero activities.
 
For cooperative rearrangements to occur, an energy barrier of height $E$ needs to be crossed, thus the relaxation time is expressed in an Arrhenius form as $\tau = f^{-1}\exp{\left(-E/k_\mathrm{B}T_\mathrm{eff}\right)}$. Here, we suppose that in the limit where $T_\mathrm{eff}$ is close to $T_0$ the extra energy provided by self-propulsion is not altering significantly the argument of the exponential. However, the attempt frequency $f$ might be altered by the process of space exploration. 
Below, we replace the many particle problem by the simpler problem of a single particle that explores a cage of size $a=0.3\sigma_0$. $f$ is then the frequency at which the test particle is coming close to the lowest barrier in the cage.

A Brownian particle explores its cage by translational diffusion in a time $\tau_\mathrm{cage}^\mathrm{B} = \mu a^2/(4k_\mathrm{B}T_0) \approx 0.1\tau_\mathrm{R}$.
Recent simulations of glassy active particles consider only a self-propulsion force, without translational diffusion~\cite{Flenner2016,Berthier2017,Nandi2018}. The persistent random walk of such a particle can be characterized by the magnitude of this force $F_\mathrm{P}$ and its persistence time, here $\tau_R/2$~\cite{Howse2007,Palacci2010}. Since the cage size is shorter than the persistent length, the elementary time of cage exploration is the persistence time, $\tau_\mathrm{cage}^\mathrm{P} = \tau_\mathrm{R}/2$.
It implies that $\tau_\mathrm{cage}^\mathrm{P}/\tau_\mathrm{cage}^\mathrm{B} = (8/3)(R_\mathrm{H}/a)^2$. This ratio depends only on the softness of the potential and is about 5 in our case. A nonBrownian self-propelled particles explores its cage five times slower than a Brownian particle.

Experimentally, our particles are submitted to both translational Brownian motion and propulsion forces. For times shorter than the persistence time, a particle thus undergoes random motion biased in the propulsion direction. This situation is analogous to the sedimentation-diffusion problem~\cite{Perrin1909}, replacing the weight by the propulsion force. Along the propulsion direction, the particle probability density follows an exponential law of characteristic length $\lambda_\mathrm{P} \equiv k_\mathrm{B}T_0/F_\mathrm{P}$, analogous to a sedimentation length. From (Eq.~\ref{eq:TeffPeH}) we get the relevant Peclet number for cage exploration
\begin{equation}
    \mathrm{Pe} \equiv \frac{a}{\lambda_\mathrm{P}} = \frac{3}{\sqrt{2}}\frac{0.3\sigma_0}{R_\mathrm{H}} \left(\frac{T_\mathrm{eff}}{T_0} -1\right)^{1/2}.
\end{equation}
The propulsion force dominates the cage exploration for $\mathrm{Pe}>1$, that occurs above the effective temperature $T_\mathrm{eff}^*/T_0 = 1 + (2/9)(R_\mathrm{H}/0.3\sigma_0)^2\approx 1.45$, that corresponds to the lowest activity we can achieve experimentally. Therefore, even at our lowest nonzero activity, diffusion is facing an uphill battle to explore the cage in the direction against propulsion. There is thus a practical discontinuity between our passive case, where the cage is explored by translational Brownian motion, and our first active case ruled by the physics of self-propelled particles. Between these two cases, the attempt frequency to cross energy barriers in the glass phase is typically reduced by a factor of 5.

To summarize, we have exhibited a dramatic change in the response of dense assemblies of colloids to low levels of self propulsion at the glass transition.
While the system is ergodic, the relaxation time decreases monotonically with activity. In the nonergodic glassy state, the relaxation time unexpectedly increases in the very first nonzero activity and then decreases at high enough activity for collective motions to kick in.
We attribute the observed slowdown to a drop in efficiency of cooperative relaxation due to the onset of directed motion and name this phenomenon ``Deadlock from the Emergence of Active Directionality'' (DEAD). 
The magnitude of the slowdown is larger than the factor of 5 found by our scaling argument taking into account space exploration of a single particle. We conjecture that the many-body nature of cooperative motion has to be taken into account to reach quantitative agreement. A reduction in attempt frequency at the single-particle scale may translate non-linearly into a larger relaxation time at the level of the cooperative region. Unfortunately recent extensions of glass theories to active matter rely explicitly on effective single-particle models~\cite{Nandi2018}.
Furthermore, we have to take into account that the number of degrees of freedom per particle jumps from 2 in the Brownian case, to 3 in the self-propelled case where orientation become important. In other words, directional motion adds $N$ degrees of orientational freedom that increase even more the complexity of the landscape and slows down relaxation. Our argument on propulsion-induced confinement shows that the switch from isotropic to oriented system is effective at very low activities.
Finally, the existence of DEAD opens the door to actively arrested materials where dynamics are even slower than in their passive counterpart.

\begin{acknowledgments}
The authors thank Ludovic Berthier, Grzegorz Szamel, Chandan Dasgupta and Takeshi Kawasaki for fruitful discussions.
N.K. is supported by PhD scholarship from the doctoral school of Physics and Astrophysics, University of Lyon.
N.K. an M.L. acknowledge funding from CNRS through PICS No 7464. 
M.L. acknowledges support from ANR grant GelBreak ANR-17-CE08-0026.
C.C.B. and C.Y. acknowledge support from ANR grant TunaMix No.
ANR-16-CE30-0028 and from Université de Lyon, within
the program Investissements d’Avenir IDEXLyon (Contract
No. ANR-16-IDEX-0005) operated by the French National
Research Agency (ANR).
\end{acknowledgments}

%


\clearpage
\newpage

\section*{Supplementary materials}

\subsection*{Experimental configuration}
Our Janus particles are made from gold particles of nominal diameter \SI{1.6}{\micro\metre} (Bio-Rad \#1652264) that we half coat with \SI{20}{\nano\metre} platinum~\cite{Howse2007}. 
The particles are dispersed in deionized water and we put them in a 96-well microplates (Falcon \#353219) where they settle down to the bottom and form a monolayer. The hydrodynamic diameter of the particle measured from the diffusion coefficient in the dilute passive case is \SI{1.9}{\micro\metre}. The plate is set on a Leica DMI 4000B microscope and we observe their 2D motion from the bottom of the plate, which is shone by a custom-made external dark-field LEDs ring. The experimental data are taken in the form of images sequence at \SI{5}{\hertz}. Particle trajectories are reconstructed by using Trackpy package~\cite{trackpy2016}.

Due to electrostatic repulsion, the particles are not at direct contact. That is why we define an effective diameter $\sigma_0$ instead of using the nominal diameter. We define $\sigma_0=\SI{2.2}{\micro\metre}$ from the position of the first peak of the radial distribution function in a dense passive regime ($\phi = 0.80 \pm 0.02$).

\subsection*{Crystalline order}
\begin{figure}
    \centering
    \includegraphics{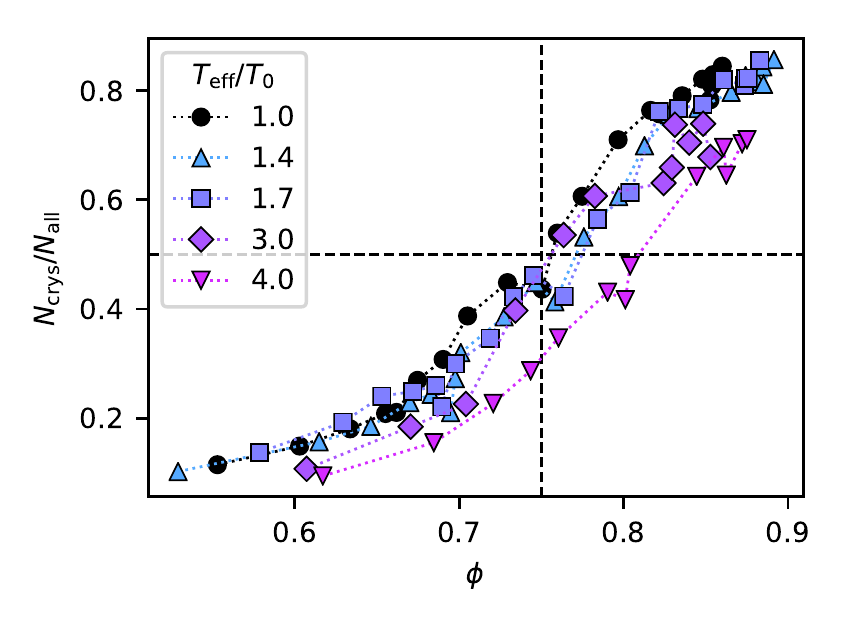}
    \caption{Density dependence of $N\rs{crys}/N\rs{all}$, a ratio of a number of crystalline particles to a total number of particles. The horizontal dashed line at  $N\rs{crys}/N\rs{all} = 0.5 $ is the threshold value where we limit our analysis in order to avoid the crystalline region. This corresponds to $\phi < 0.75$ (the vertical dashed line).}
    \label{fig:crys}
\end{figure}

\begin{figure*}
    \centering
    \includegraphics{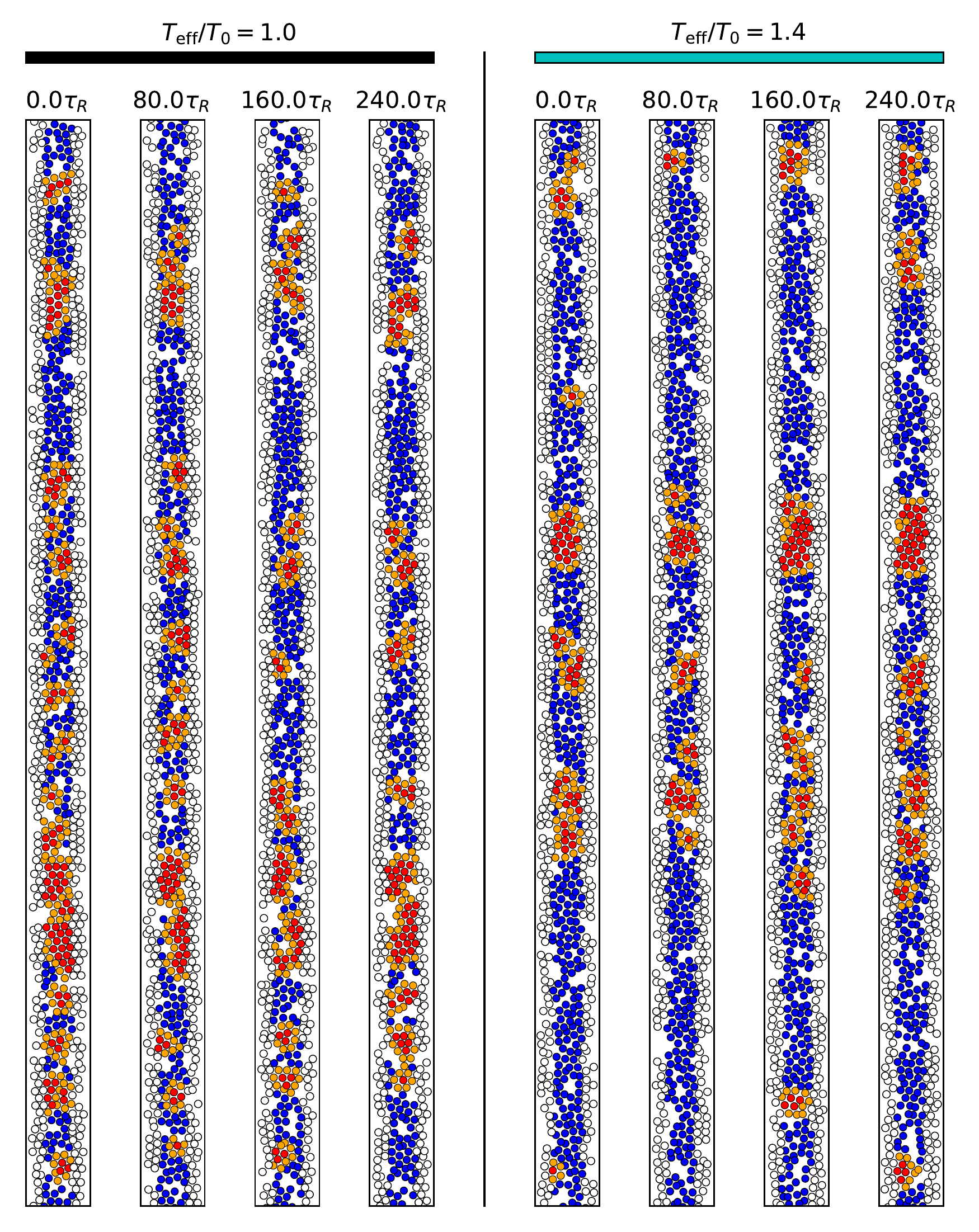}
    \caption{Snapshots at several times of a slice at $\phi = 0.72$ in passive (left) and the first nonzero active (right) cases. Particles are displayed with different colors: Crystalline particles (red), particles in direct contact with crystalline particles (orange), particles nearby the edges (white), and the remaining particles (blue).}
    \label{fig:map_glass_den72}
\end{figure*}

Our Janus particles always contain doublets and triplets. However these defects do not totally prevent crystal nucleation, especially at high densities. The degree of local ordering is quantified using the hexatic order parameter~\cite{Nelson1979}, 
\[
\psi_{6,i} = \frac{1}{6}\sum_{j\in n_i} \exp(6\imath\theta_{i,j})
\]
where $n_i$ is a set of 6-nearest neighbors of a particle $i$ and $\theta_{i,j}$ is an angle between particles $i$ and $j$. Here we calculate a ratio of crystalline particles (particles that have $|\psi_{6,i}| > 0.8$) to all particles, $N\rs{crys}/N\rs{all}$, at various activity levels and a range of density. Fig.~\ref{fig:crys} shows that the system order decreases when the activity level is higher. Fig.~\ref{fig:map_glass_den72} displays the spatial arrangement of crystalline regions that are rather compact and stable in time, with a visible decrease of local ordering with increase of activity.

\begin{figure}
    \centering
    \includegraphics{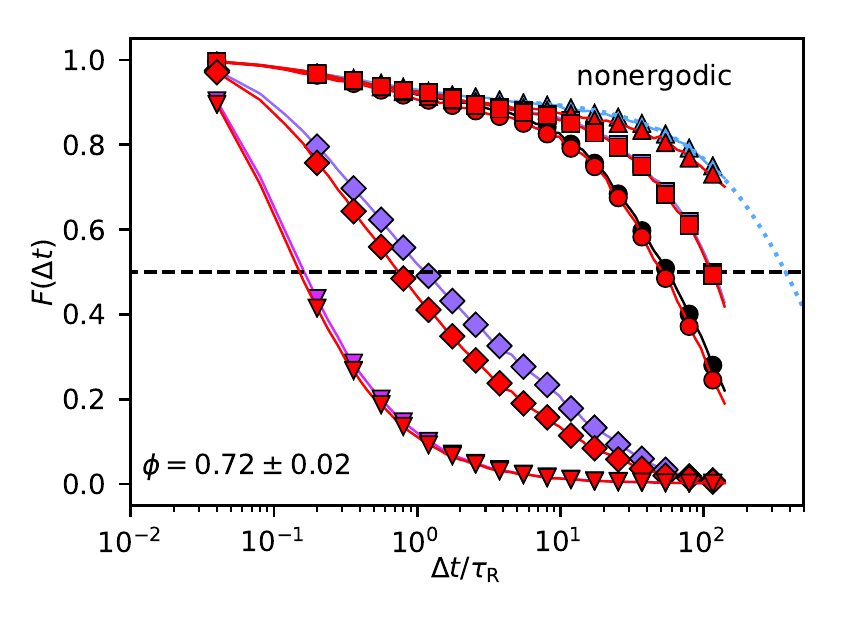}
    \caption{Same $F(\Delta t)$ plot as in Fig. 1 of the letter, with (in red) the neighbours of crystalline particles excluded from the calculation.}
    \label{fig:Ft_Xngb}
\end{figure}

High values of $N\rs{crys}/N\rs{all}$ at large $\phi$ reveals that the lowest part of the sediment is practically a polycrystal.
Therefore, in this letter we restrict our analysis to densities where $N\rs{crys}/N\rs{all} < 0.5$ at any activity, which corresponds to $\phi < 0.75$. Furthermore, we explicitly exclude the trajectories of crystalline particles out of our computations of $F(\Delta t)$, and thus from the definition of the relaxation time $\tau$. Nevertheless, including or excluding crystalline particles, or even their neighbours, has little influence on $F(\Delta t)$, as shown in Fig.~\ref{fig:Ft_Xngb}. Moreover, our companion paper shows that, in the densest part of the sedimentation ($\phi > 0.80 \pm 0.02$) where the system is mostly crystalline, we  still observe the DEAD behavior.

\subsection*{Broken bonds}
One way to investigate the relaxation mechanisms of the system is by considering whether a particle keep or change its 6-nearest neighbors. Two particles are said to be neighbors if they both belong to each-others 6-nearest neighbors and the distance between them is less than $1.5\sigma_0$. This neighborhood relation defines the edges of a graph where the particles are the nodes. Using \textit{NetworkX} package~\cite{networkX} we analyse the time evolution of this graph.
If two particles where initially neighbours but no more after a lag time $\Delta t$ we define a broken bond.
Cooperative relaxation in a passive glass usually involves diffuse broken bonds. Collective motion, on the other hand, involves more broken bonds localized at the boundary between correlated domain.
Here at the slice of $\phi = 0.72 \pm 0.02$ we compute bond correlation, which is the ratio of bonds that remain unbroken after a lag time $\Delta t$.

\begin{figure}
    \centering
    \includegraphics{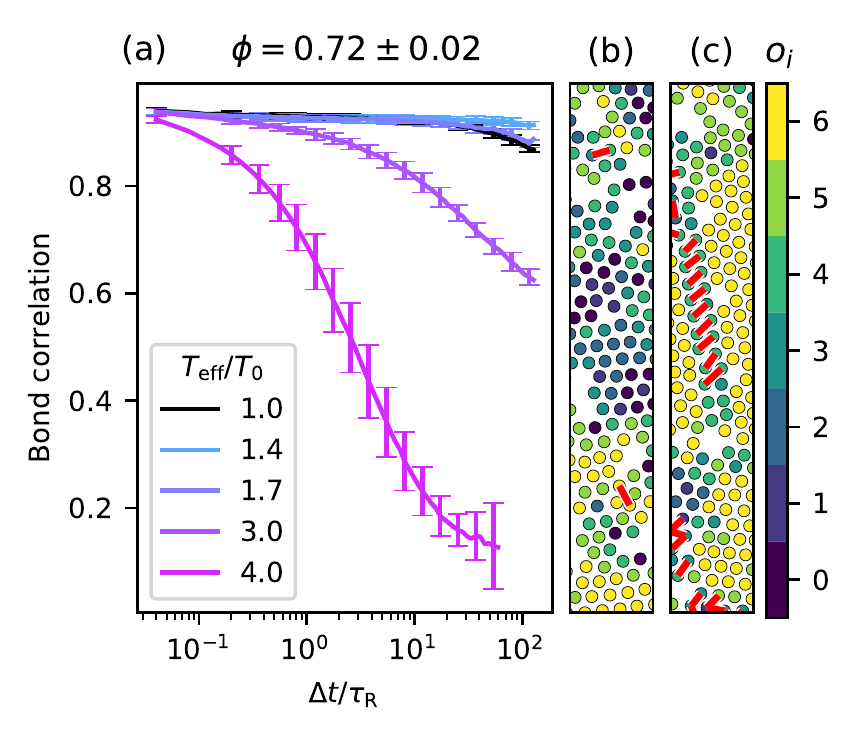}
    \caption{(a) Bond correlation with $\Delta t$ at $\phi = 0.72 \pm 0.02$ and various activities. (b,c) Directional correlation maps, $o_i$, for $T\rs{eff}/T_0 = 1.0$ and $3.0$, respectively, at $\Delta t = 12\tau\rs{R}$. The red lines between two particles represent broken bonds during $\Delta t$. The white areas are from sample artifacts and tracking errors.}
    \label{fig:broken_bonds}
\end{figure}

Fig.~\ref{fig:broken_bonds}a shows the bond correlation at various activity levels.
We can notice that the bonds relaxation follows the ``back and forth'' behavior.
Furthermore, the bonds relaxation is clearly seen at higher activities where the system has entered the liquid phase (see Fig.~\ref{fig:diagram}b in the main text). 

In Fig.~\ref{fig:broken_bonds}b,c we show the directional correlation maps, $o_i$, in the passive case and $T\rs{eff}/T_0 = 3.0$, respectively. $o_i$ is defined as
\begin{equation}
    o_i = \sum_j \Theta\left(\frac{\vec{u}_i \cdot \vec{u}_j}{|\vec{u}_i||\vec{u}_j|} - 0.5\right),
\end{equation}
where $\vec{u}_i$ is the displacement of particle $i$ in a given lag time $\Delta t$ and $\Theta$ is the Heaviside step function. In other words, $o_i$ is a number of neighbors of particle $i$ (among six neighbors) that have the same orientation of displacement as $i$.
The broken bonds are presented by red lines between two particles. In the passive case, the broken bonds are sparse and not localized (see Fig.~\ref{fig:broken_bonds}b) showing the cooperative relaxation mechanism.
At $T\rs{eff}/T_0 = 3.0$ the system is in the liquid phase, but it is still close to the glass transition. We are thus able to observe broken bonds at the boundary between two groups of particles travelling in opposite directions (see Fig.~\ref{fig:broken_bonds}c). This is a clear signature of collective motion.

\subsection*{Waiting time dependence}

\begin{figure}
    \centering
    \includegraphics{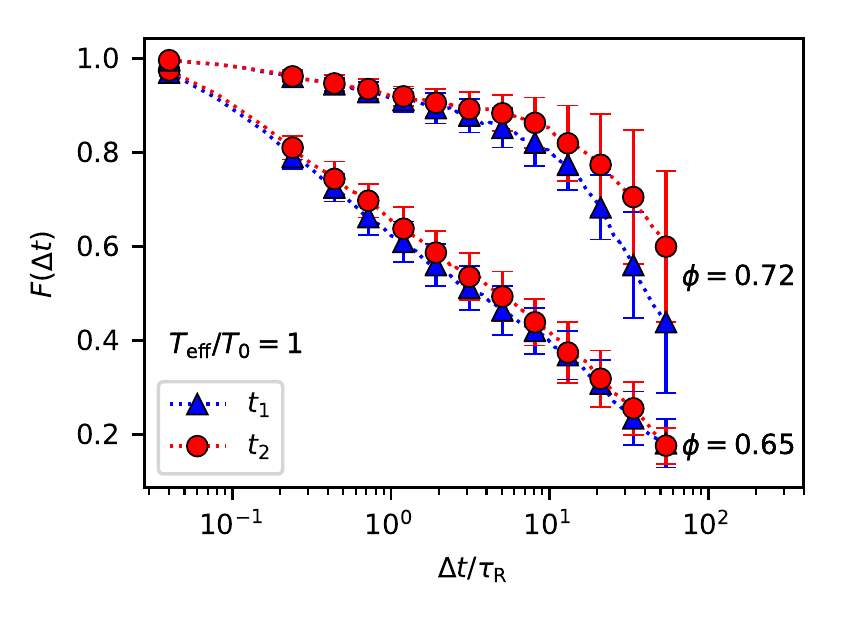}
    \caption{Relaxation function $F(\Delta t)$ of the passive case at different waiting times $t_1$ and $t_2$ such that $t_2 - t_1 = 140\tau\rs{R}$. The result is compared between ergodic ($\phi = 0.65 \pm 0.02$) and nonergodic state ($\phi = 0.72 \pm 0.02$). The time unit is divided by Brownian reorientation time $\tau\rs{R} = \SI{5}{\second}$.}
    \label{fig:aging}
\end{figure}

In an ergodic system, the relaxation function $F(\Delta t)$ should depend only on the lag time $\Delta t$, not on the reference time. By contrast, the relaxation of a nonergodic system depends on the waiting time. To confirm that we cross the ergodicity limit between $\phi = 0.65 \pm 0.02$ and $\phi = 0.72 \pm 0.02$, we investigate the waiting time dependence of $F(\Delta t)$ in the passive case. We select two waiting time separated by $140\tau\rs{R}$ and show the result in Fig.~\ref{fig:aging}.

At $\phi = 0.65 \pm 0.02$ there is no significant difference between the two waiting times. We thus confirm that at this density the system is ergodic. It is where we found that the relaxation time responses monotonically to activity level. 
At $\phi = 0.72 \pm 0.02$ where we observe the nonmonotonic behavior, the relaxation function depends on the waiting time. We thus confirm that the system is nonergodic at this density. We thus show that the nonmonotonic behavior appears when crossing the ergodic to nonergodic limit. 

The data acquisition in the active case are always performed at the same waiting time: $360\tau\rs{R}$ after new \ce{H2O2} was added to the system.

\end{document}